# MODELING AND SIMULATION OF DATA PROTECTION SYSTEMS FOR BUSINESS CONTINUITY AND DISASTER RECOVERY

Sašo Nikolovski [1] and Pece Mitrevski [2]

[1] Faculty of Informatics, AUE-FON University, Skopje, North Macedonia
[2] Faculty of ICT, University "St. Kliment Ohridski", Bitola, North Macedonia

## ABSTRACT

*In today's corporate landscape, particularly where operations rely heavily on information technologies, establishing a robust business continuity plan, including a disaster recovery strategy, is essential for ensuring swift recuperation following outages. This study presents a comparative analysis of recovery solutions, focusing on systems that operate partially or entirely within cloud environments and assessing their reliability in fulfilling organizational roles securely and dependably. Two such systems were deployed and evaluated in a real-world production setting. Key performance and reliability metrics were identified using simulation software to enhance these systems, alongside a System Dynamics analysis conducted for each. This work proposes a comprehensive framework for selecting and maintaining data protection and recovery solutions within organizational structures, outlining criteria for aligning chosen approaches with operational needs while adhering to predetermined timelines specified in business continuity and disaster recovery plans. The resulting analysis and findings offer actionable insights to guide decision-making when selecting appropriate recovery concepts.*

## 1. INTRODUCTION

The constant availability of company information systems for many imposes an impression and opinion that business continuity can be interpreted or understood as protecting the future business and functional survival of the organization from some form of disruption. In contemporary business environments that rely on digital systems, achieving *zero downtime* during operational disruptions is regarded as the optimal goal for organizations committed to continuity. Meeting this expectation is not always feasible or practical due to various factors, such as weather-related outages or cyber attacks. While organizations have access to a range of protection and recovery solutions, both within their local data centers and through cloud-based systems [1], certain disruptions may be unavoidable. Consequently, from an organizational management perspective, there is a heightened focus on minimising the effects of outages on overall operations by determining the maximum allowable outage duration that can be sustained without causing lasting consequences for future activities.

When evaluating outages and establishing goals for reliable recovery that meet organizational standards, it's important to recognize that the process mainly revolves around how long the organization remains non-operational. Additionally, this process involves two distinct time-based factors: one is determined by the system or technological aspect presented by the Recovery Time

Objective (RTO), and the other, which is more organizationally oriented, represents the time required for a full operational recovery of work processes (Work Recovery Time-WRT).These two time-components of the recovery process determine the Maximum Tolerable Downtime (MTD) provided by the Business Continuity Plan (BCP) and the Disaster Recovery Plan (DRP) and is the summary time of the recovery time objective and the required operational time return to work processes, i.e., MTD = RTO + WRT.

This means that RTO as a parameter is a time interval during which operations are performed in the technical-technological part of the organization, a time during which systems, data and network infrastructure are restored. The remaining time until the maximum tolerable outage time is the operational recovery time (WRT) and in it is carried out recovery of all work processes that are based on information and information systems (

Figure 1). The recovery time frame, which is limited by the framework in which the MTD is set, also includes several components that are an integral part of this process. These components aim to restore data from their backup which is closest in time to the outage, to carry out subsequent operations as the final part of the recovery and, of course, to check and test the functionality of the systems before they are officially operational to establish normal work in organization.

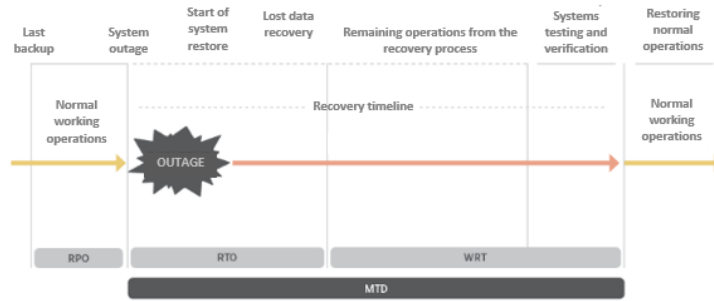

Figure 1. Maximum Tolerable Downtime (MTD)

When defining the needs, in the system design phase, the component that is directly related to the data, and thus to the consistency of the information, is the Recovery Point Objective (RPO). This parameter is time-dependent and gives the "age" of the data in the backup set at the point in time from which it will be restored to the systems for operational use. Considering that the RPO recovery process is backward recovery, loss of data and final information is inevitable (except in synchronous replication situations where data loss is zero). Therefore, when designing systems within the framework of sustainable business continuity planning, the maximum tolerance threshold is set for data loss that the organization can afford (Maximum Tolerable Data Loss-MTDL). Systems that enable zero data loss create an inversely proportional relationship between the amount of data lost during the recovery process and the cost of the systems. This means that the closer the RPO is to the time of the outage, the higher the cost of such systems and vice versa.

For that reason, using the above-mentioned timecomponents as a starting point, an analysis of the performance and reliability of two systems for recovery after an outage is carried out, with the aim of obtaining a parametric framework for the selection and installation of systems that will satisfy the requirements of organizations during recovery processes from a technical-technological, organizational and financial aspect, which very often has a decisive influence when making the final choice.

The rest of the paper is organized as follows.Section 2 provides an overview of prior research on solutions pertaining to cloud services as well as approaches addressing disaster recovery processes as a conceptual framework. The third section describes practical solutions implemented for recovery from outages and disasters, outlining the necessary requirements, the structural components used, and their specific features within these solutions. Two recovery system scenarios are under consideration. The first is a hybrid approach, featuring an on-premises data center solution that integrates cloud-based security as an endpoint. The second scenario involves a fully cloud-based recovery system. Section 4 outlines the identification and definition of key parameters used for the analysis and evaluation of the proposed systems. These parameters are functionally linked through the application of the System Dynamics approach to each system individually, followed by a comparative analysis. Conducting "what-if" simulations to test specific policies on such a model significantly enhances our understanding of how the system evolves, helping to identify which parameters should be targeted in recovery plans and ensuring business continuity within organizations. During the summary of the results and their graphical representation in Section 5, the performance of the components involved in the data protection process is concurrently evaluated. This approach aims to provide a comprehensive understanding of system performance within data center security operations. Section 6 presents the concluding observations derived from the existing disaster recovery systems and the evaluation of their performance. It emphasises the primary strengths and limitations of each option within the context of their respective application scenarios.

## 2. RELATED WORK

This study draws upon a series of publications generated from research conducted over the past decade, a period characterized by the widespread adoption of cloud services within routine business operations [2][3][4][5][6][7][8][9][10]. In the context of these papers, several key parameters are routinely considered; among them, the Recovery Point Objective (RPO) and Recovery Time Objective (RTO) are frequently highlighted. These parameters are directly linked to the performance and reliability of data protection systems [11]. It is important to note that the majority of these studies were conducted under simulation conditions, typically utilising key parameter values derived from isolated environments without the impact of other infrastructure components. Consequently, the authors emphasise that their findings require validation in an actual production setting.

Mitrevski *et al.* [2] have reviewed a plethora of works that have been published over a decade, contributing in invaluable ways to the area of cloud performance and dependability modeling. Their focus is on the use of a class of Stochastic Petri Nets (SPNs) with reward structures as an integrated part of modeling, known as Stochastic Reward Nets (SRNs). In addition, they have proposed a framework for performability modeling of a cloud service by applying SPNs and discrete-event simulation (DES) for the evaluation of the corresponding metrics.

In [3], a review and description of the parameters by which such solutions are evaluated is made with additional analysis of specific techniques for DR with their advantages and disadvantages. The research prioritises the techniques and principles underlying BC/DR provision. The evaluation of each concept involves a thorough analysis using several criteria, including economic cost-effectiveness, privacy protection, feasibility of implementation, and most importantly, reliability, which serves as the cornerstone for all other considerations. Tabular reviews present final ratings for factors such as safety, redundancy, complexity, and recovery time; however, they do not detail RPOs or RTOs in relation to data size or system volume. Including this information would provide more precise and comprehensive insights in both the tables and accompanying text.

Within [6], a comparison was made between traditional DR solutions based on *hot site / cold site* and a solution based on Disaster Recovery as a Service – DRaaS. Rebah and Sta analyze the importance of disaster recovery planning (DRP) as a critical component of business continuity planning (BCP), driven by organizations' need to safeguard data structures during information system outages. Their research presents a comparative analysis of current disaster recovery solutions, with particular emphasis on evaluating scenario-based approaches to disaster recovery as a service in cloud environments. Additionally, the research solely concentrates on analysing and evaluating the performance achieved through cloud services. It also specifically references studies by Gartner, Forrester Consulting, and Aberdeen Group regarding the proportion of these services used in disaster recovery solutions. Considering the potential risks associated with these services, the authors reference study [12] conducted by Lexsi, France's first cybersecurity service company. This research identifies eight primary risks that users should carefully evaluate before deciding to implement DRP solutions in the cloud. When selecting DRaaS, certain factors warrant close attention, yet there are no recommendations provided on whether such solutions are suitable or unsuitable for small and medium-sized businesses. In the conclusion, the authors pose a significant unanswered question: is entrusting my data to a third party truly an effective way to ensure its protection? This dilemma is highly relevant in today's systems and plays a pivotal role in choosing a cloud service, potentially outweighing other previously considered criteria.

Mendonca *et al.* in a couple of their research papers, have made a serious approach to the research of systems and services for DR, of which in [8] they give a special reference to analysis and modeling for the assessment of Backup as a Service-BaaS. This research relies on analytical models and outage experiments to assess crucial disaster recovery parameters, including Recovery Time Objective (RTO), Recovery Point Objective (RPO), availability, and downtime duration. Continuing their work, Mendonca et al. in [9] investigate the availability of a disaster recovery solution using multiple criteria. Similar to [8], they employ DSPN networks for modeling but also introduce the Multiple-Criteria Decision-Making (MCDM) method to evaluate and rank various disaster recovery solutions.

A different DR strategy is discussed in [13], focusing on cloud environments with single-cloud and multi-cloud scenarios. Key parameters include Critical Business Function (CBF), Maximum Acceptable Outage (MAO), RTO, and Business Impact Analysis (BIA), along with their interdependencies. The authors note that these solutions often result in longer recovery times, leading to higher RTO and Cost of Downtime (CoD).

A comprehensive analysis of a cloud-based service solution is presented in [14], evaluating multiple scenarios by comparing cost and performance metrics for both cloud-based and on-premises environments. One notable limitation is the omission of Recovery Time Objective (RTO) values in each scenario, which also precludes the calculation of Cost of Downtime (CoD). Furthermore, the text does not specify the volume of data protected against loss in either scenario. The authors in [15] present a detailed description and methodology based on Business Impact Analysis (BIA), with particular emphasis on key considerations for designing and constructing an IT Disaster Recovery (DR) system. Through implementation, they successfully address the main limitation of the existing infrastructure – downtime – with results demonstrating a reduction of this parameter by up to 85%. However, a notable limitation is the lack of even a basic system description, which would provide greater clarity regarding the type of DR system discussed. Additionally, there is an absence of value analysis for key parameters essential to such solutions, including RPO, RTO, and CoD. Furthermore, critical infrastructure data, such as the number of systems requiring protection, the capacity and integrity of storage systems, and the methods used to connect to the global network for maintaining continuous network services, are also not sufficiently evaluated.

In practical terms, all these authors note that most methods and algorithms in their research are developed and tested under simulated conditions, where key parameter values are applied in isolation without environmental influences. To confirm their real-world applicability, these approaches require validation through practical application.

## 3. Methodology

Thesystems used in this research are hosted in a production data center(

Figure 2).

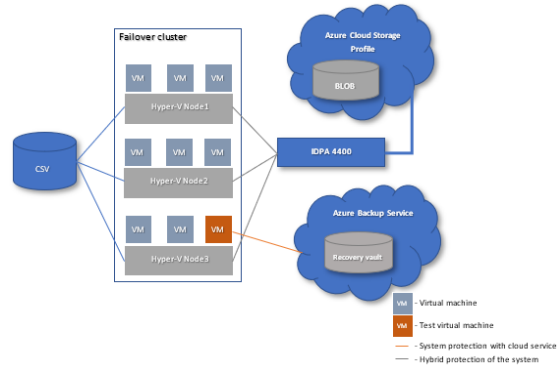

Figure 2.  Production data center–block diagram

Considering that the research is conducted in an actual working environment where the stable functionality of all systems is of great significance, a dedicated server system was established as a test virtual machine (VM) for the research. This VM was utilized to oversee the processes related to data protection (backup) and recovery (recover/restore) within the context of the two protection systems, which consist of protective storages located in the cloud.In order to determine the performability of data protection systems, within the framework of the research, two basic and two extended models were developed in which, by simulating the operation of real systems, values for the target parameters that are of interest for drawing conclusions were obtained. The practical implementation of the systems used in the research includes the application of a hybrid solution (a hardware device with a data storage layer placed in the cloud) placed in the data center, with an Avamar software agent installed in the server system that is subject to data protection and a solution based entirely on on the Microsoft Azure Recovery Service (MARS) cloud technology[16], with the MARS agent installed in the server system. When backing up data using the hybrid solution, the backup is initially stored in the device's local storage for a pre-set period of time, after which the copy is transferred to the cloud storage tier for long-termkeeping. In the case of backups using the MARS agent, the backups are uploaded and stored directly in cloud storage, with a preset retention period.

The overall process included a series of activities consisting of:

1.  Creating a detailed project, setting up and configuring a data center including data protection based on a DellEMC DP4400 hardware device [17][18][19]as an on-premises solution and Microsoft Azure Recovery Services (MARS) as a fully cloud-based solution,
2.  Installing and configuring data protection systems for research purposes. The number of samples of time and data parameters retrieved from the systems was matched with the values predicted by the business impact analysis (BIA) and the minimum number of samples supported by the protection systems (14 days from the hybrid system and 7 from the cloud-based system),

3. After a time period of one year in which the systems had continuous operation, an analysis of their operation was performed by downloading the data on time and data parameters from the performance of backup and recovery operations) of the data, where the parameters that should be calculated from the values of the parameters taken from the protection systems are determined,

4. For each of the considered systems, two models have been created – onebasic model, in which, by using the values taken from the systems for the time and data parameters, the values for the determined derived parameters in the model have been obtained, and one extended model, in which, by using the values of the derived parameters from the basic model are calculated values for the performance of the systems with a given test amount of data in the considered operating environment. Also, during the creation of the models, a model for the reliability of the systems was created, with time frames aligned with used time settings made in the systems,

5. A comparative analysis of the obtained results from the simulations of the models for the two systems was made and conclusions were drawn based on which recommendations for correct dimensioning and placement of the different protection concepts in different situations of use were given.

According to everything stated above, the focus of the research is set in the area of maintaining business continuity in the operations of organization entities, and the analysis of the reliability and performance of systems for disaster recovery is imposed as a subject of research, while the goal is set to building a parametric framework that will provide precise guidelines when choosing a system solution for the protection and maintenance of data and information systems in data centers.

## 4. MODELS, SIMULATIONS AND RESULTS

When setting up the models for the protection systems used in the research, the main emphasis in them is placed on the time to perform backup operation and time to restore/recover data. What is significant about the developed models is that they use values taken from the real systems as a source for the input values of the variables. The initial values for data quantities are taken from the agents, for a time sample of 14 days in the case of the hybrid system and 7 days in the case of the cloud-based system, in order to cover the expiration of the copies, but also to show the relationship of the repositories in cloud with the conditions set in the data center for the transfer of data to them. The business impact analysis of outages (BIA), which is part of data center protection policies, serves as the foundation for determining the parameters used when configuring the models. The target values established by the BIA are listed in Table 1, where they are separated according to specific parameters for each system individually.

The table shows that for hybrid systems, the Avamar agent is installed on the server being backed up, while the MARS agent is used for cloud systems.

Table 1. Parameters values in BIA

| Agent | Backup frequency | Backup retention time | Recovery points in time | Cloud tiering policy | RPO | RTO |
|-------|-----------------|----------------------|------------------------|---------------------|-----|-----|
| Avamar | daily | 14 | 7+7+60 | > 14 days | ≤ 7 days | ≤ 5 hours |
| MARS | daily | 7 | 2*7+3 | / | | |

The backup scenario is fully compliant with the values given in Table 1.To monitor and evaluate the process of recovering lost data from a selected recovery point in time (RPO), a simulation of damage (deletion) of document folders is made, on which the recovery process has been carried out.

Figure 3 shows, in a general form, the input-output parameters in the models of the two systems considered.

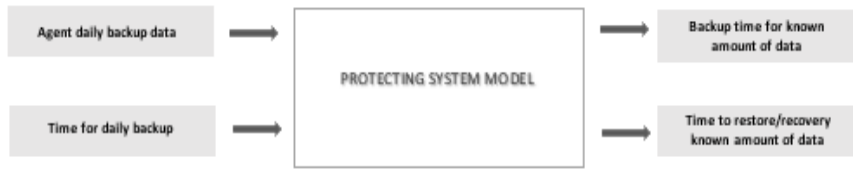

Figure 3.Input-output parameters in the models

From

Figure 3, it can be noted that the amount of data transferred on a daily level to the storage systems and the time for which they were transferred to the storage systems, are set as input parameters for both models. Derived parameters for the time needed to protect a given amount of data and the time for its recovery within the system that is subject to protection are defined as output parameters from the models.

During the development of models for both systems, the foundational models are first established. These encompass a set of variables that are interrelated, allowing them to influence the states of resulting components and storage systems within their respective frameworks.To evaluate system performability during data protection and recovery processes, an extended model was developed for each system. These models incorporate additional components that enable the measurement of specific time parameters associated with backup and recovery operations.

## 4.1. Hybrid System Model

The basic model of the hybrid system is shown in Figure 4, where there are four derived components that cover the process of making backup copies of the data, as well as the process of recovery in case of data damage or lose. The time frame in which the simulation process takes place in the given model is aligned with the BIA, where the process of creating backup copies takes place in 14 time periods, and the process of recovering the data and moving it to the cloud storage level is set in one period after the backup time frame.

Values of the variables in the basic model are shown in

Table 2 with separate views of the components of the backup process, the restore process and values of the derived variables in the model.

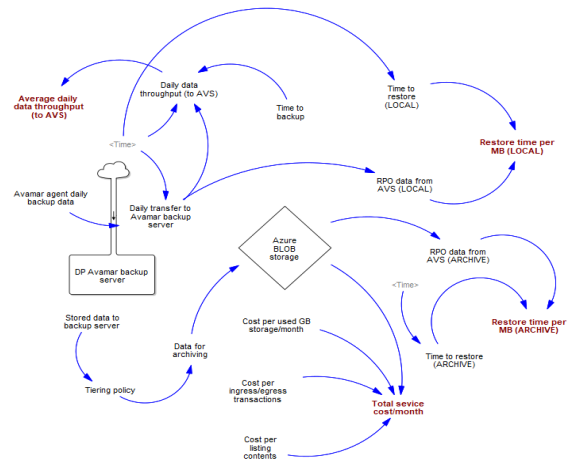

Figure 4. Basic model of a hybrid system with data tier in cloud

In Figure 4, *Average daily data throughput (to AVS)* appears as derived component in the model, a derived variable obtained as a mean value of the amount of data transferred to the protection storage in the system (Avamar backup server - AVS) by the Avamar agent in the server system, *Restore time per MB (LOCAL)* variable which shows time required to restore a 1MB amount of data retrieved from the local storage of the DP4400 system, *Restore time per MB (ARCHIVE)* variable which shows the time required to restore a 1MB amount of data that was moved to the storage level of the device placed in the cloud and of course the derived variable *Total service cost/month* which shows the costs of using a storage service placed in the cloud.

Figure 5 provides a graphical states representation of input components in the model to present parameter values changes inside of the specified time frame.

Table 2. Values of variables set in the basic model

| Variable | Value | | | | | | | | | | | | | | |
| --- | --- | --- | --- | --- | --- | --- | --- | --- | --- | --- | --- | --- | --- | --- | --- |
| | 1 | 2 | 3 | 4 | 5 | 6 | 7 | 8 | 9 | 10 | 11 | 12 | 13 | 14 | 15 |
| **Data backup process** | | | | | | | | | | | | | | | |
| Avamar agent daily backup data (MB) | 26956 | 25712 | 27194 | 26710 | 25147 | 24520 | 26529 | 19711 | 27342 | 19574 | 27024 | 25262 | 24644 | 26082 | |
| Daily transfer to Avamar backup server (MB) | 26956 | 25712 | 27194 | 26710 | 25147 | 24520 | 26529 | 19711 | 27342 | 19574 | 27024 | 25262 | 24644 | 26082 | |
| Daily data throughput (to AVS) (MB) | 51.3448 | 40.3022 | 49.8059 | 51.8661 | 55.7633 | 57.5587 | 45.7428 | 74.6629 | 43.0596 | 76.1693 | 47.9149 | 58.3472 | 55.132 | 51.4438 | |
| Time to backup (minutes) | 8.75 | 10.633 | 9.1 | 8.583 | 7.516 | 7.1 | 9.666 | 4.4 | 10.583 | 4.283 | 9.4 | 7.216 | 7.45 | 8.45 | |
| **Data restore process** | | | | | | | | | | | | | | | |
| RPO data from AVS (ARCHIVE) (MB) | | | | | | | | | | | | | | | 1824 |
| RPO data from AVS (LOCAL) (MB) | | | | | | | | | | | | | | 1824.01 | |
| Time to restore (ARCHIVE) (seconds) | | | | | | | | | | | | | | | 470.1 |
| Time to restore (LOCAL) (seconds) | | | | | | | | | | | | | | 38.24 | |
| **Derived variables** | | | | | | | | | | | | | | | |
| Average daily data throughput (to AVS) | | | | | | | | | | | | | | | 54.2224 |
| Restore time per MB (LOCAL) (seconds) | | | | | | | | | | | | | | | 0.0209649 |
| Restore time per MB (ARCHIVE) (seconds) | | | | | | | | | | | | | | | 0.25773 |
| Total sevice cost/month (dollars) | | | | | | | | | | | | | | | 1.57912 |

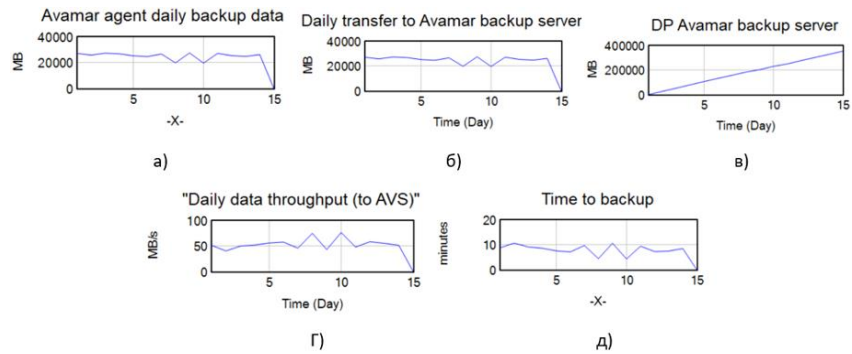

Figure 5.Graphic representation of parameters' states in the backup process

According to the subject and objectives of the research to determine the performability of the solution, the basic model of the system has been upgraded with five new components that have no influence on the values in the basic model (
Figure 6).

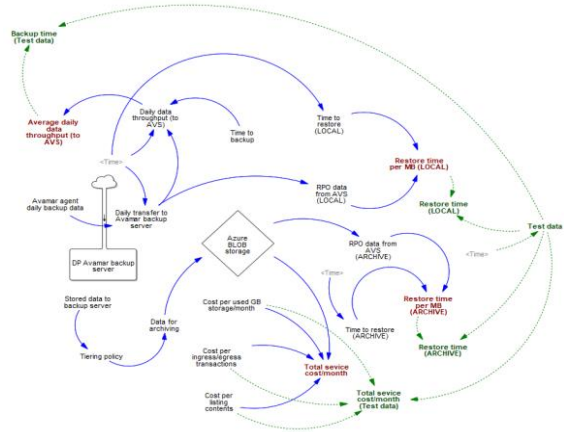

Figure 6. Extended model of a hybrid system with data tier in cloud

From the presented view of the extended model, the relationships between the derived components of the basic model and the additional components for a given test amount of data in the extended model can be observed. As additional components in the model, the *Test data* component that enters the test amount of data (531 GB), *Backup time (Test data)* as a component that calculates the time required to make a backup copy of a given amount of data, *Restore time components (LOCAL)* and *Restore time (ARCHIVE)* in which the time needed to recover a given amount of data is calculated in the case when the process is performed from the local storage of the system with RPO ≤ 14 days or from the level of storage placed in the cloud where 15 ≤ RPO ≤ 60 days and the component *Total service cost/month (Test data)* which gives the total monthly costs for using the storage service of the cloud-based system.

The values of the additional components set in the expanded model of the system are shown in Table 3, from where it can be noted that some of the resulting components in the expanded model repeatedly exceed the maximum allowed values for them provided for in the BIA, which makes them useless for the needs of the organization.

Specifically, the *Restore time from (ARCHIVE)* component has a value many times higher (38 hours) than the maximum value provided in the BIA (≤5 hours) and therefore, the level set in the cloud, in this case could not be used for quick recovery of systems in the organization with acceptable downtime and returning its operation and functionality to the level before the occurrence of the outage. Against this value, the value of the *Restore time (LOCAL)* component is within the limits predicted by the BIA (3 hours and 5 minutes), which indicates that the system placement in the data center structure can satisfy the requirements seted in the analysis.

Table 3. Simulation results of an extended model for a hybrid system

| Variable | Value |
|---|---|
| **Derived variables (basic model)** | |
| Average daily data throughput (to AVS) (MB/s) | 54.2224 |
| Restore time per MB (ARCHIVE) (seconds) | 0.25773 |
| Restore time per MB (LOCAL) (seconds) | 0.0209649 |
| **Test data simulation values (extended model)** | |
| Test data (MB) | 531012 |
| Backup time (Test data) (hours) | 2.72034 |
| Restore time (ARCHIVE) (hours) | 38.016 |
| Restore time (LOCAL) (hours) | 3.09239 |
| Total sevice cost/month (Test data) (dollars) | 11.6602 |

From the point of view of creating a backup copy, the values of the *Backup time (Test data)* component fully satisfy the requirements for quick creation of the copies in the terms provided for their implementation.

It should be considered that such systems have a complex architecture not only in hardware, but also in software, where through a series of algorithms over time of use, the time for which the copies will be made can be drastically shorter than what was taken as the average time in the basic model.

## 4.2. Cloud System Model

The basic model of the cloud-based system is shown in

Figure 7, where, just like in the hybrid system, four derived components appear, two of which relate to the process of creating a backup copy, one to the process of data recovery, and one to the monthly costs for using such a service completely set up in the cloud.

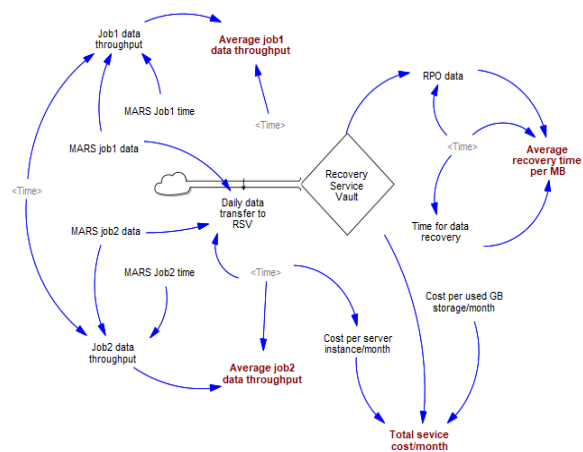

Figure 7.Basic model of a cloud-based system

The presentation of the components from which the model is built is made in the same way as the hybrid solution was described. *Average job1 data throughput* and *Average job2 data throughput* appear as derived components in the model, as variables that show the average value of the amount of data transferred to the system's backup storage (Recovery service vault - RSV), the variable *Average recovery time per MB* that shows the time required to recover 1MB amount of data placed in the protective storage of the system and *Total service cost/month* as a derived variable that shows costs of using the storage service placed in the cloud.

The time period in which the procedures for making copies and restoring data from them are performed includes 8 time points, 7 of which are intended for performing the policy for making backup copies, and the last time point is intended for showing the change in quantity data that are placed in the recovery vault after the completion of the 7-day cycle, according to the values given in Table 1. The states of the values of the variables in the basic model and their changes in the set time frame are shown in Table 4, where a separate section also shows the final values of the derived components after a completed simulation with the given input parameters.

Figure 8provides a graphical representation of the states of the input components in the model associated with the *MARS job1* process, and

Figure 9 provides a graphical representation of the states of the components associated with the *MARS job2* process.

Table 4. Value states of the variables in the basic model

| Variable | Value | | | | | | | |
|---|---|---|---|---|---|---|---|---|
| | **1** | **2** | **3** | **4** | **5** | **6** | **7** | **8** |
| **Data backup process** | | | | | | | | |
| MARS job1 data (MB) | 8362 | 8409 | 8463 | 8516 | 8569 | 8622 | 8678 | |
| MARS Job1 time (sec) | 3270 | 2993 | 3025 | 3110 | 2976 | 3105 | 3307 | |
| Job1 data throughput (MB/s) | 2.55719 | 2.80956 | 2.79769 | 2.73826 | 2.87937 | 2.77681 | 2.62413 | |
| MARS job2 data (MB) | 353 | 375 | 478 | 553 | 551 | 394 | 780 | |
| MARS Job2 time (sec) | 351 | 365 | 489 | 628 | 232 | 358 | 387 | |
| Job2 data throughput (MB/s) | 1.0057 | 1.0274 | 0.977505 | 0.880573 | 2.375 | 1.10056 | 2.0155 | |
| Daily data transfer to RSV (MB) | 8715 | 8784 | 8941 | 9069 | 9120 | 9016 | 9458 | |
| **Data recovery process** | | | | | | | | |
| RPO data (MB) | | | | | | | | 7690 |
| Time for data recovery (sec) | | | | | | | | 1380 |
| **Derived variables** | | | | | | | | |
| Average job1 data throughput (MB/s) | | | | | | | | 2.57731 |
| Average job2 data throughput (MB/s) | | | | | | | | 1.83045 |
| Average recovery time per MB (sec) | | | | | | | | 5.57246 |
| Total sevice cost/month(dollars) | | | | | | | | 7.82701 |

The simulation results related to the derived components listed in Table 4 provide the foundation for evaluating the system's performability, especially when manipulating extreme values of data components that directly affect system performance.

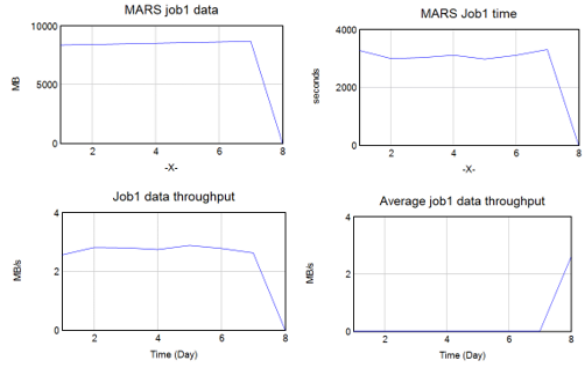

Figure 8. Graphical representation of component values related to MARS job1

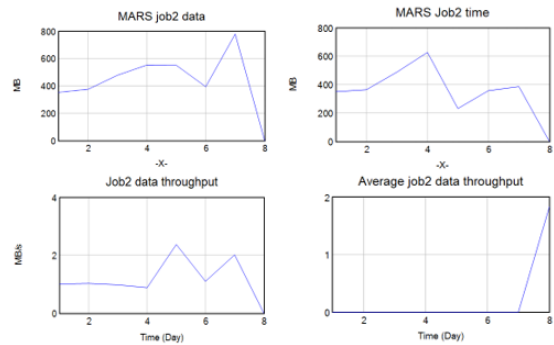

Figure 9.Graphical representation of component values related to MARS job2

For this purpose, five new components have been added to the model that is presented as basic and from which the derived components are used, of which:

- Two components*(Backup time Job1(Test data)*and*Backup time Job2(Test data))*for time calculationin backup creation with both jobs,
- a component that will refer to the data recovery process *(Recovery time (Test data))*,
- *Test data*is a common component for all, which will carry the value of the amount of data that a server system has in the organization (Test data = 531 GB) and
- one component *(Total service cost/month (Test data))*for calculating the costs of using the service with the new amount of data.

The view of the expanded model with the new components is given in
Figure 10, and the results of the simulation are shown in
Table 5.

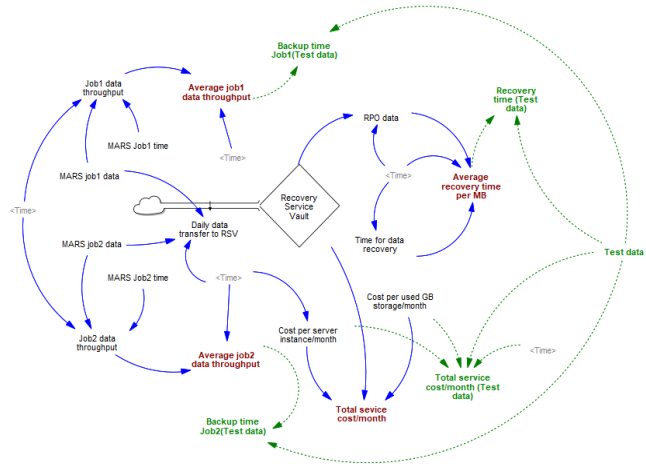

Figure 10.Extended model of the cloud-based system

Table 5.  Simulation results of an extended model for the cloud-based system

| Variable | Value |
|---|---|
| **Derived variables (basic model)** | |
| Average job1 data throughput (MB/s) | 2.57731 |
| Average job2 data throughput (MB/s) | 1.83045 |
| Average recovery time per MB (sec) | 5.57246 |
| **Test data simulation values (extended model)** | |
| Test data (MB) | 531012 |
| Backup time Job1(Test data) (hours) | 57.2315 |
| Backup time Job2(Test data) (hours) | 80.5831 |
| Recovery time (Test data) (hours) | 26.47 |
| Total sevice cost/month (Test data) (dollars) | 43.7893 |

Due to the high values of the time components, representation of these values in
Table 5is given in hours, and the data values in MB.

## 5. COMPARATIVE ANALYSIS

In order to obtain a complete picture of the performability of the two systems, a comparative
analysis of the results obtained from the performed simulations of the presented models was
made. By applying the values obtained from the data protection systems, in a comparative
overview divided by operations in three tables, components that are marked as performed in the
simulations are placed. In the first table (Table 6) a comparative overview of the amounts of data
transfer in both systems during the process of creating a backup copy of the data is made.

Table 6.  Values of data transfer when creating a backup copy

| SYSTEM | Hybrid (DP 4400) | CLOUD-BASED (Azure recovery service) | |
|---|---|---|---|
| Component | Average daily data throughput (to AVS) | Average job1 data throughput | Average job2 data throughput |
| Value (MB/s) | 54.2224 | 2.57731 | 1.83045 |

The review indicates that the hybrid system exhibits significantly higher data transfer rates between the Avamar agent and the Avamar server (AVS) during backup creation. This increased throughput is primarily attributed to the communication pathway: in the hybrid system, connectivity occurs via the local network, whereas in the cloud-based system the process originates with the MARS agent in the data center, traverses the Internet service provider (ISP), and concludes at the cloud-based service. Changes to data transfers reduce performance, resulting in lower total bandwidth and longer copy creation times. Table 7 presents a comparison of simulation results for data recovery processes and their derived components in both systems.

Table 7. Time component values in recovery process for 1MB data

| SYSTEM | Hybrid (DP 4400) | | CLOUD-BASED (Azure recovery service) |
|---|---|---|---|
| Component | Restore time per MB (LOCAL) | Restore time per MB (ARCHIVE) | Average recovery time per MB |
| Value (sec) | 0.02096 | 0.25773 | 5.57246 |

The data presented in the table indicate that the hybrid system enables significantly faster recovery of data from the device's local storage compared to processes that retrieve data from cloud storage.

The use of cloud storage by the described data protection and recovery systems also entails costs that result from the use of the cloud service. Table 8 shows these costs and the parameters according to which they were made.

Table 8. Monthly costs for using the cloud service for both systems

| SYSTEM | Hybrid (DP 4400) | CLOUD-BASED (Azure recovery service) |
|---|---|---|
| Cost per server instance/month | / | 10 |
| Cost per used GB storage/month | 0.02 | 0.0448 |
| Cost per ingress/egress transactions ($/10K operations) | 0.54 | / |
| Cost per listing contents ($/10K operations) | 0.5 | / |
| TOTAL COST for TEST DATA ($/month) | 11.66 | 43.79 |

The last row in the table shows the cost of using storage space equal to the amount of data (531 GB) that was used in the simulations for both systems in the cases of their extended models.

## 5.1. Reliability Analysis

According to the layout of the components in the system that shows the concept of data protection from outages and disasters, it includes three components: the data center (DC), the cloud service (Azure) and the connection to the global network (ISP). With this concept of the system, it can be shown as a series connection between its components where the failure of any component in the system (not the operational binary state of the component, for which R=0 will apply), will mean the failure of the entire system. This way of thinking about the operation of the entire system is the basis for analyzing its reliability. When building a model for calculating and assessing the reliability of the system, in our case it implies setting more variables in it that will represent the conditions on which the reliability of each of its components depends. The initial conditions for setting up the concept of the model include: the period in which the data center will be used (predicted 7 years or 61320 hours), the period in which the system analysis is performed (15 days or 360 hours), the operational time of the connection operator to the global network (based on the usage contract of 24 months or 17520 hours) and the level of Service Level Agreement (SLA) of 99.95%.

The reliability analysis model of data protection concepts with the initial conditions set in this way is shown in
Figure 11, and the results of the performed reliability simulation of each of the components and the system as a whole are given in Table 9.

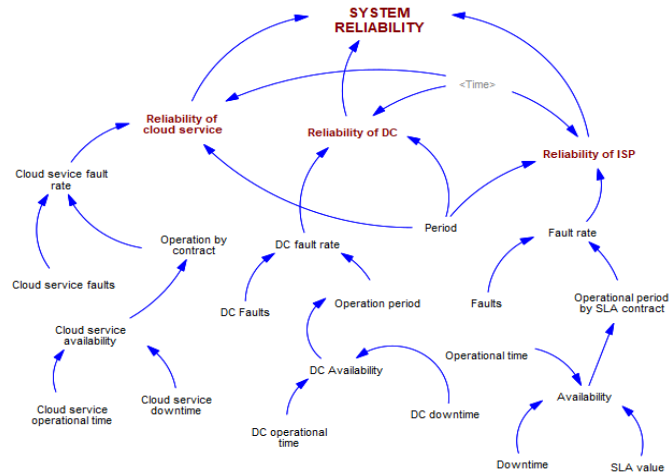

Figure 11. A three-component system reliability model

From the results presented in Table 9, it can be concluded that despite the high reliability values of the cloud service and the data center (reliability for them in the initial conditions is set to R=1), the lower reliability value of the operator has a negative impact on the reliability of the whole system. It is due to the serial connection of these components in the execution of the processes that use the cloud service. In the Disaster Recovery Journal (DRJ) [20], it is stated that any scenario or concept of such a solution is as reliable as its connection to the global network. Therefore, in concepts of systems that are based on a cloud service or a service of any kind that uses a connection to the global network, when designing such solutions, a careful assessment of the values in the SLA documents is mandatory, because small deviations in them contribute for noticeable changes in the characteristics of the rest of the components, and thus of the entire set solution.

Table 9. Reliability of components and system for a given period

| COMPONENT | VALUE |
| --- | --- |
| Reliability of cloud service | 0.993952 |
| Reliability of DC | 0.993952 |
| Reliability of ISP | 0.978981 |
| SYSTEM RELIABILITY | 0.967174 |

## 6. CONCLUSION

Within contemporary business operations and the rapid pace of digital transformation, the significance of data and the systems housing it has become increasingly apparent, positioning data as the central element in information exchange.Business processes increasingly rely on protective solutions to ensure uninterrupted operations. Numerous systems exist for this purpose, each offering distinct benefits and drawbacks regarding how they safeguard business continuity. Some focus specifically on secure data storage through backups, while others protect active data by keeping copies at multiple locations with ongoing or periodic synchronization (replication). Additionally, certain solutions are designed to maintain several organizational data centers, further supporting continuous business activity without interruption. The key contribution of this paper is identifying parameters essential for selecting solutions that meet BIA requirements and ensure business continuity in organizations reliant on information systems. Accordingly, this paper presents concluding observations categorized by technical aspects, characteristics and performance, as well as financial considerations. The final section provides insights and recommendations intended to guide future research aimed at enhancing performability and reliability in data protection systems. Recognizing that data volume directly affects the timing components of protection solutions, future research should prioritize mechanisms that minimize the amount of data exchanged between systems throughout the protection process. From both a technical perspective and based on how well protection solutions work, results from models developed for two systems show that placing a local solution which distributes data to the cloud as part of the device's overall storage (using a single data namespace) offers adequate protection during outages. This approach ensures a fast recovery to an operational state for information and systems needed for daily activities. When evaluating the costs associated with cloud services, it is important to recognize that the primary factor increasing the *Total service cost/month* is the charge for storage space usage, particularly when storing large volumes of data. The total amount of data stored directly influences this component of the overall cost.

## AUTHORS

**Sašo Nikolovski** received his BSc in Informatics from from the "Ss. Cyril and Methodius" University in Skopje, MSc degree in MIS, and PhD degree in Informatics and Computer Technology, respectively, from the University "St. Kliment Ohridski" – Bitola, North Macedonia. He is currently an adjunct assistant professor at the Faculty of Informatics, American University of Europe – FON, Skopje, North Macedonia. As a system administrator, he has more than 22 years' experience in designing hardware infrastructures, data protection and disaster recovery solutions based on physical and virtual appliances from the Dell Technologies portfolio, designing and maintaining logical infrastructures based on Windows and VMware technologies and design, implementation and administration of complex network infrastructures.

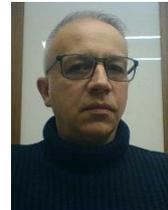

**Pece Mitrevski** received his BSc and MSc degrees in Electrical and ComputerEngineering, and the PhD degree in Computer Science from the "Ss. Cyril and Methodius" University in Skopje, North Macedonia. He is currently a full professor at the Faculty of Information and Communication Technologies, University "St. Kliment Ohridski" – Bitola, North Macedonia anda lecturerat the Department of Computer Sciences, Faculty of Economics, Technology and Innovation,Western Balkans University – Tirana, Albania. His research interests include Computer Architecture, Computer Networks, Cloud Computing, Performance and Reliability Analysis of Computer Systems, e-Commerce, e-Government, e-Learning, and Information Security. He has authored over 130 publications in reputable journals and refereed conference proceedings, and has delivered numerous lectures on these subjects. He holds membership in both the IEEE Computer Society and the ACM.

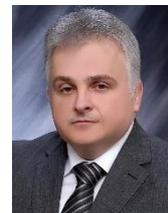